\documentclass[sigconf]{acmart}

\AtBeginDocument{%
  \providecommand\BibTeX{{%
    \normalfont B\kern-0.5em{\scshape i\kern-0.25em b}\kern-0.8em\TeX}}}
\setcopyright{ccbysa} 
\copyrightyear{2022}
\acmYear{2022}

\acmConference[Intelligent Music Interfaces CHI`22 Workshop]{CHI Conference on Human
Factors in Computing Systems}{May 01, 2022}{New Orleans, LA, USA}
\usepackage{todonotes} 
\usepackage{caption}
\usepackage{subcaption}
\begin{document}

\title{Proper Posture: Designing Posture Feedback Across Musical Instruments}

\author{Bettina Eska}
\orcid{}
\affiliation{%
  \institution{LMU Munich}
  \city{Munich}
  \postcode{80337}
  \country{Germany}}
\email{bettina.eska@ifi.lmu.de}

\author{Jasmin Niess}
\orcid{0000-0003-3529-0653}
\affiliation{%
  \institution{University of St. Gallen}
  \city{St. Gallen}
  \country{Switzerland}
}
\email{jasmin.niess@unisg.ch}

\author{Florian Müller}
\orcid{0000-0002-9621-6214}
\affiliation{%
  \institution{LMU Munich}
  \streetaddress{Frauenlobstrasse 7a}
  \city{Munich}
  \country{Germany}
  \postcode{80337}
}
\email{florian.mueller@ifi.lmu.de}

\renewcommand{\shortauthors}{Eska et al.}

\begin{abstract}
There is a recommended body posture and hand position for playing every musical instrument, allowing efficient and quick movements without blockage. Due to humans' limited cognitive capabilities, they struggle to concentrate on several things simultaneously and thus sometimes lose the correct position while playing their instrument. Incorrect positions when playing an instrument can lead to injuries and movement disorders in the long run. Previous work in HCI mainly focused on developing systems to assist in learning an instrument. However, the design space for posture correction when playing a musical instrument has not yet been explored. In this position paper, we present our vision of providing subtle vibrotactile or thermal feedback to guide the focus of attention back to the correct posture when playing a musical instrument. We discuss our concept with a focus on motion recognition and feedback modalities.
Finally, we outline the next steps for future research.
\end{abstract}



\maketitle

\section{Introduction}
Learning to play a musical instrument requires time and many hours of practice to overcome the difficulties in the beginning. Learners have to concentrate on controlling the musical instrument, playing the correct rhythm, and reading the note sheet simultaneously. Due to human's limited cognitive capacities, they cannot keep their attention focused on several things over a longer period of time \cite{paas1994cognitiveLoad}. This can lead to the loss of the correct body posture or hand position when the musician concentrates intensely on the musical text. However, maintaining the proper posture is essential when playing a musical instrument. For every musical instrument, there is a recommended body posture that, for example, facilitates the control of airflow and breathing in wind instruments or allows hands and fingers to move quickly without any blockage. According to \citet{metcalf2014complexHandDex}, there is a relationship between finger velocity or hand dexterity and skilled performance on a musical instrument. Maintaining good posture can make finger movements more efficient and help to achieve the correct timing and precision~\cite{goebl2013tempControlHandMove}. By not correcting incorrect positions, the musician becomes accustomed to them, which makes the incorrect posture even more difficult to correct in the long term \cite{masaki2012CognitiveNeuroscience}. Furthermore, incorrect postures when playing a musical instrument can lead to injuries or movement disorders in the long run, especially when practicing extensively \cite{furuya2013flexibilityPiano}. Consequently, incorrect posture when playing an instrument is problematic for beginners who should not acquire bad habits and advanced musicians who often invest a considerable amount of time playing their instrument.

There exists a large body of related work on feedback systems that support the learning of motor skills for playing a musical instrument using a variety of sensory output modalities. 
Within the context of musical instrument learning, research proposed various systems using visual feedback. For instance, \citet{rogers2014PIANO} project the music notation onto the piano for faster learning. Like this, pianists learn the correct fingering through different color highlights. As another example, \citet{marky2021letsFrets} propose to support guitar learning by visually highlighting the target fret on the guitar neck. While the presented solutions support the musician in the general learning process for their specific instrument, they do not provide feedback on the posture. In the related field of dance learning, \citet{anderson2013youmove} and \citet{hulsmann2019superimposed} provided an augmented skeleton as an overlay on virtual mirrors. They found that users were able to reduce performance error by comparing their own performance with the optimum.
While practical and useful, the visual sense is often occupied in motor learning; therefore, other modalities or multi-modal feedback can enhance motor skill learning of complex tasks~\cite{sigrist2015sonification}.

As another popular output modality, research also explored auditory feedback to support motor learning and found that auditory feedback is especially suitable for overcoming difficulties with rhythm by enhancing the beat~\cite{jylha2011auditoryRhythmic}. 
Auditory feedback is not suitable for practice in groups because it interferes with the music and, thus, can be disturbing for listeners, for example, in concerts or other musicians because of contradictory information. Therefore, it is only suitable for individual practice at home.

To overcome the limitations of visual and auditory cues, research explored haptic feedback modalities that are only noticeable for the musician, mostly unobtrusive, and do not disturb others. In research, the term \textit{haptic} refers to both tactile (e.g., vibration, pressure, etc.) and kinesthetic (e.g., muscle receptors, body pose, etc.) feedback~\cite{Sigrist2013AugmentedMultimodalFeedback}
Previous work showed that tactile feedback is effective for posture corrections~\cite{rotella2012hapi, spelmezan2009tactile} and indicating directions~\cite{hong2017evaluatingHapticFeedb} by guiding the focus of attention~\cite{Grosshauer2009augmentedHaptics}. Users can discriminate different tactile patterns with high accuracy in physical activities. As a first step towards supporting musicians maintaining the correct pose while playing instruments, \citet{Grosshauer2009augmentedHaptics} integrated a vibrotactile actuator in the violin's bow to provide tactile feedback on the bow stroke, which had a positive impact on the performance. Exploring temperature as another haptic output modality, \citet{Peiris2019ThermalBracelet} showed the potential of spatiotemporal thermal stimuli in a guidance scenario. Users could detect the direction of the thermal stimuli on the wrist with high accuracy. However, to the best of our knowledge, there is no prior work on using temperature feedback to support users in keeping the correct hand posture when playing instruments.

Considering the respective advantages and disadvantages of the discussed output modalities, in this work, we argue that haptic feedback systems are the more suitable solution to support musicians. Such feedback systems can provide subtle nudges to direct the focus of attention to support the learner in real-time~\cite{Grosshauer2009augmentedHaptics} without interfering with the primary task of playing the instrument, which already loads the visual and auditive channels of the user. The feedback should not distract the musician, nor should it require cognitive effort to interpret it. Furthermore, referring to the example of a pianist, such a system could be used to train the optimal hand position that allows fast finger movements while protecting the joints. As recent research explored ways to measure the body position, in the remainder of this paper, we focus on the feedback by providing subtle nudges to guide the musicians back to the correct posture without distracting them.

\section{Vision}
\begin{figure}
         \centering
         \includegraphics[width=0.4\textwidth]{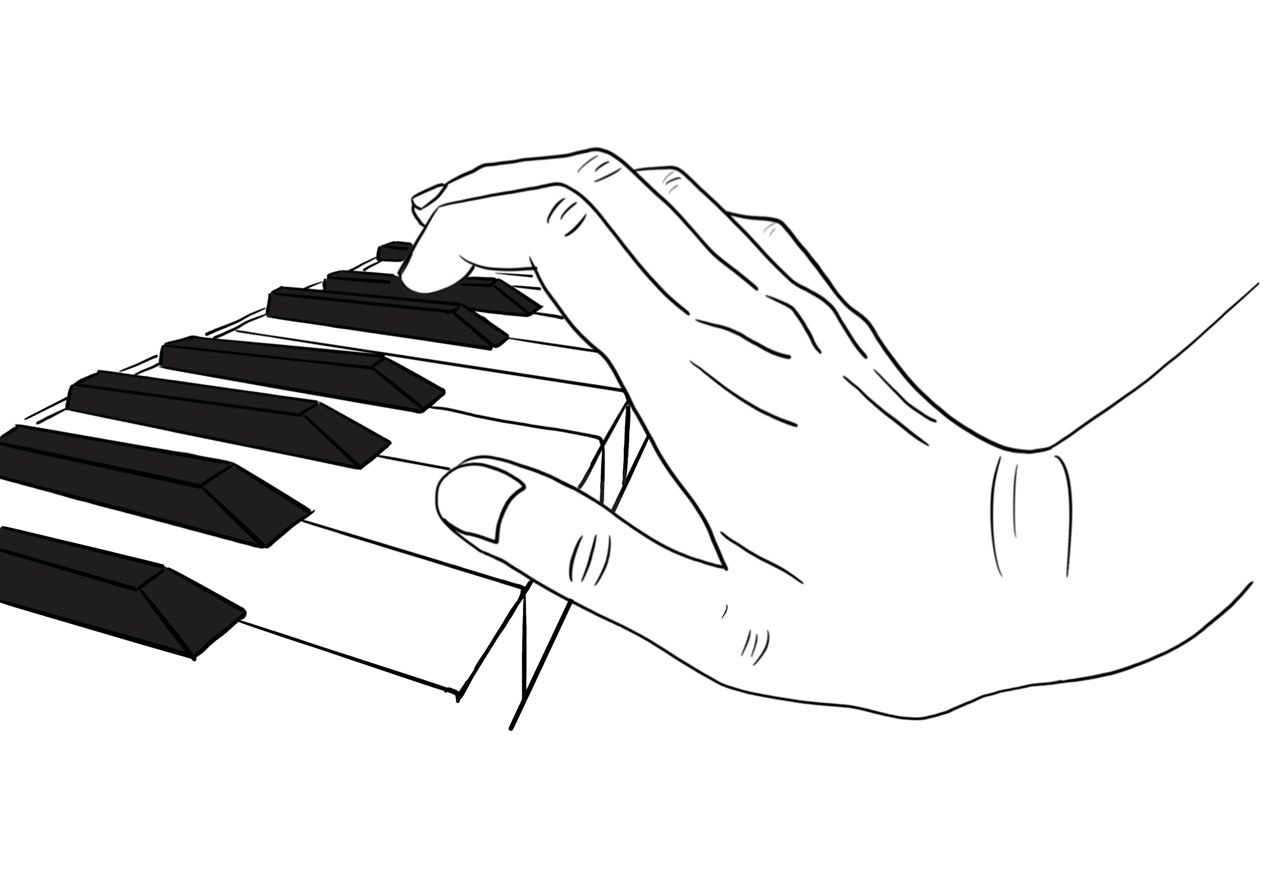}
         \caption{Incorrect hand position when playing the piano}
         \label{fig:incorrect}
\end{figure}
Building on recent findings, our vision is to explore the design space of haptic feedback systems to support musicians in keeping the correct posture using the piano as an example. We propose a system that, using the piano as an example, detects the incorrect posture of musicians during playing their instrument and generates haptic feedback based on the deviation. We envision the system to automatically track the position and posture of the users' wrist similar to systems to measure tendon loads of guitar players~\cite{sung2013developmentBiomechanical} or finger position on string and keyboard instruments~\cite{Grosshauser2013FingerPos}. Based on the calculated deviation to the correct position, the system will provide subtle feedback through haptic channels to nudge the player into the correct hand position.

We chose the wrist position of pianists as the example use case. Figure~\ref{fig:incorrect} shows a hand position that the pianist should avoid.
\subsection{Motion recognition}
For such a system, it is necessary to identify the correct position for the targeted instrument. To establish a proper baseline, we plan to recruit experts on the instrument to provide their theoretical knowledge and generate sample data sets by motion-recording their play. Common technologies for movement measurement are motion-capture systems, electromyography (EMG), or Inertial Measuring Units (IMUs) to record acceleration and gyroscope data~\cite{furuya2013flexibilityPiano, Grosshauer2009augmentedHaptics} combined with different processing and data analysis methods and machine learning to interpret the data. To support the musician during practice, we must ensure that the system calculates the deviation quickly to provide meaningful feedback in real-time. 

\subsection{Feedback}
The system should provide feedback based on the deviation of the current execution from a pre-defined optimum as identified in the data gathering with expert players. The system can convey the feedback to the musician through different sensory modalities, e.g., auditory, visual, and haptic. As discussed above, the provided feedback must be noticeable and intuitively understandable, and at the same time, non-intrusive. It is not necessary to provide feedback if pianists maintain an incorrect position for a couple of seconds or while switching to a different octave on the keyboard, but correct them if they maintain an incorrect posture for a longer time, e.g., over multiple piano keys. 
We want to avoid continuous feedback on the body posture because, as known from other motor learning contexts, it can distract the learner and become annoying~\cite{Raheb2019DanceInteractiveLarning, Sigrist2013AugmentedMultimodalFeedback}.

In this position paper, we focus on haptic, namely vibrotactile and thermal feedback. The system we suggest should recognize when the musician keeps an incorrect position over a certain time and give feedback by nudging them. The feedback system should generate haptic patterns from the wrist over the back of the hand to guide the musician unobtrusively towards the correct position during playing. We plan to explore vibrotactile and thermal feedback as options for posture correction. Both modalities are already used to create the illusion of a moving stimulation on the skin~\cite{lee2012saltation, liu2021thermocaress}. Research found that with vibrotactile patterns, the users perceived a pull metaphor instead of a push metaphor as more natural to recognize the indicated direction~\cite{camarillo2018guidance, spelmezan2009tactile}. With vibrotactile feedback, we can provide highly accurate and efficient feedback at discrete points of the skin. However, vibrotactile feedback can also interfere with the actual task of playing because it can sometimes still be heard~\cite{wilson2011SomeLike} and might constitute too much of an interference in the musician's playing process. Thermal feedback, in contrast, is entirely private and only noticeable for the user~\cite{wilson2011SomeLike}. The downside of thermal feedback is that it has a higher ambiguity than vibrotactile feedback and the response time is lower~\cite{narumi2009DesignThermal}. Nevertheless, we propose to explore thermal feedback in comparison to vibrotactile feedback because we want to assess its suitability for subtle nudges without interfering with the main task.

\begin{figure}
     \centering
     \begin{subfigure}[b]{0.4\textwidth}
         \centering
         \includegraphics[width=\textwidth]{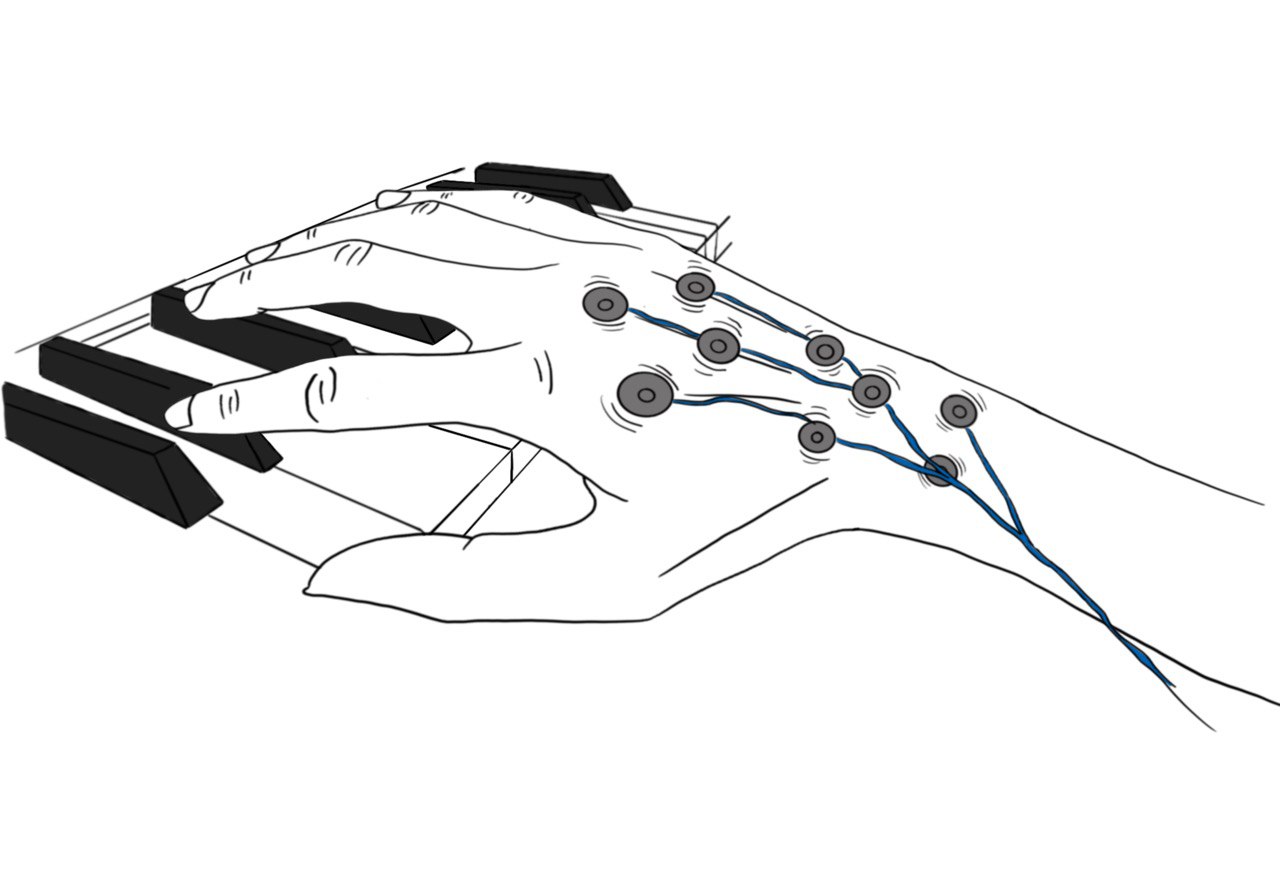}
         \caption{Vibrotactile feedback to correct the hand position}
         \label{fig:vibro}
     \end{subfigure}
     \hfill
     \begin{subfigure}[b]{0.4\textwidth}
         \centering
         \includegraphics[width=\textwidth]{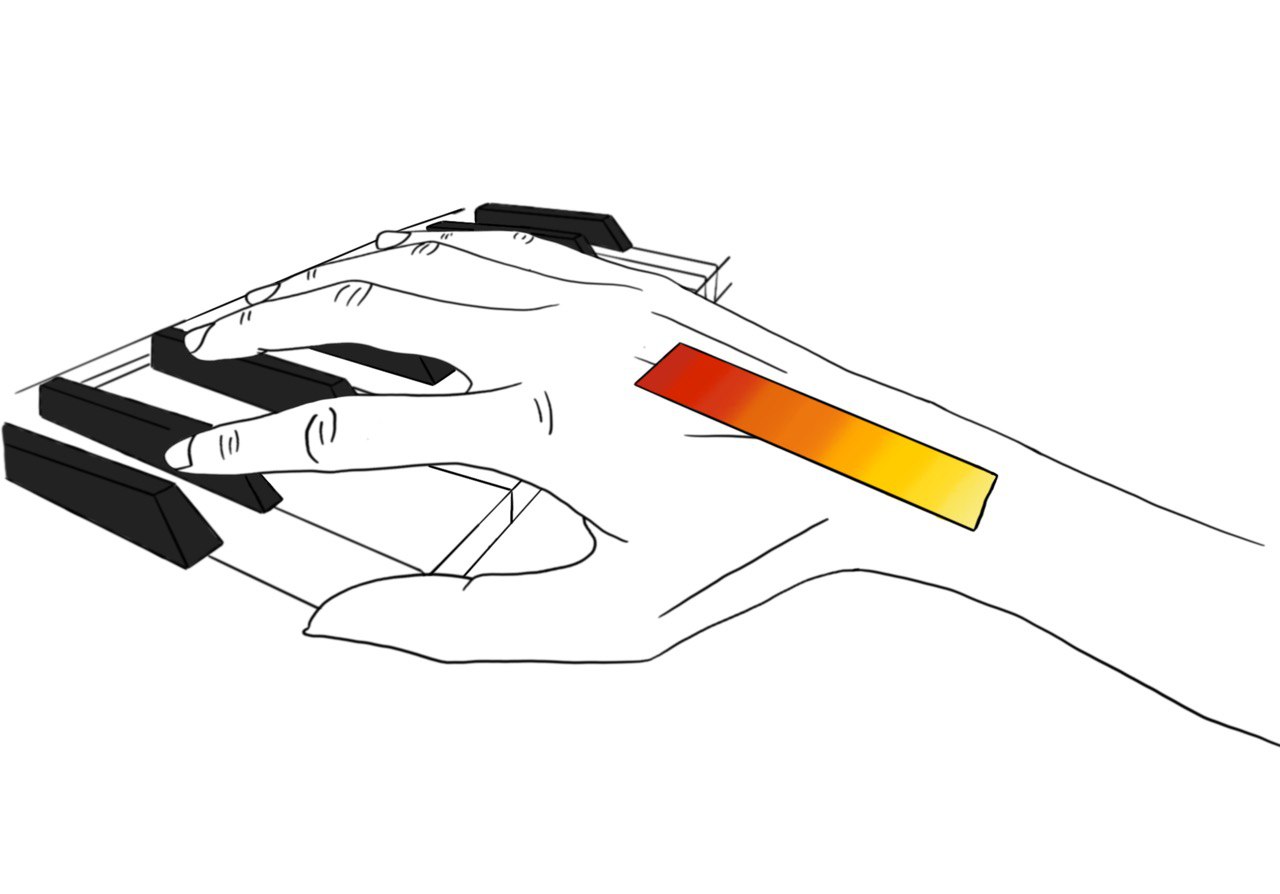}
         \caption{Thermal feedback to correct the hand position}
         \label{fig:thermal}
     \end{subfigure}
	 \caption{Feedback system to support musicians to keep the correct hand position}
     \label{fig:system}
\end{figure}
Figure~\ref{fig:system} shows a possible feedback system attached to the back of the musician's hand. Whereas Figure~\ref{fig:vibro} shows a system using vibrotactile feedback, and Figure~\ref{fig:thermal} uses thermal feedback. When the system detects a deviation from the optimum, such as a too low wrist position, it activates the feedback. For the vibrotactile feedback, this means that the vibration motors start vibrating one after the other, starting at the wrist and, thus, guiding the musician with the pull metaphor to lift the wrist again. 
When using thermal feedback, the temperature of thermal elements increases gradually, starting with a low temperature on the wrist and reaching the maximum on the back of the hand. As researchers have already studied the push and pull metaphor of vibrotactile feedback, we want to focus on which information is transferable with thermal feedback. 
The main focus of the planned system is to provide the feedback in a way that it is subtle and not distracting, and it does not require cognitive effort to interpret it, thus not interrupting the flow when playing the instrument.


\section{Discussion \& Conclusion}
In this position paper, we shared our vision of designing a feedback system that supports musicians with subtle haptic feedback to keep the correct hand position during their play. We proposed a system design consisting of a posture tracking system combined with haptic feedback to nudge the player towards the correct position. To achieve this, we envision a system that identifies the correct posture per instrument, e.g., the wrist posture of the pianist, and tracks the position. When the position deviates for several bars from the optimum, the musician gets feedback that nudges him towards the correct position. We assume that musicians internalize the position when they are directed back again and again by the system. Then no more effort is needed, and they can hold it automatically. Such a system could be an essential support tool for beginners and advanced learners. Every musician should get accustomed to the correct posture from the beginning or sometimes needs to be reminded of it, as this will prevent injuries and movement disorders from extensive practice in the long run.
Future studies can target the question of the feasibility and applicability of haptic feedback systems to support practicing a musical instrument. 
Furthermore, we plan to evaluate whether vibrotactile or thermal feedback is easier to notice and understand when playing a musical instrument and perceived as less annoying.
In this position paper, we used the piano as the target instrument, but we are confident that the concept is transferable to other instruments. Given that data about correct body posture exists, the system can provide feedback for musicians playing the guitar, violin, flute, or other instruments. Further, the concept is also applicable beyond the context of musical instruments. For example, a posture evaluation system providing haptic nudges can remind the user to keep an upright position when sitting on a chair.


\begin{acks}
Thanks to Anna Scheidle for helping with the figures.
\end{acks}

\bibliographystyle{ACM-Reference-Format}
\bibliography{proper-posture}

\appendix

\end{document}